\documentclass[twocolumn,letterpaper]{IEEEAerospaceCLS}  % only supports two-column, letterpaper format

\usepackage[T1]{fontenc}
\usepackage[latin9]{inputenc}
\usepackage{geometry}
\usepackage{float}
\usepackage{amsmath}
\usepackage{amsthm}
\usepackage{amssymb}
\usepackage{graphicx}
\usepackage{float}
\usepackage{balance}

\makeatletter

\floatstyle{ruled}
\newfloat{algorithm}{tbp}{loa}
\providecommand{\algorithmname}{Algorithm}
\floatname{algorithm}{\protect\algorithmname}

\theoremstyle{plain}
\newtheorem{thm}{\protect\theoremname}
\theoremstyle{plain}
\newtheorem{lem}[thm]{\protect\lemmaname}

\usepackage{algorithm,algpseudocode}
\algnewcommand{\Inputs}[1]{%
  \State \textbf{Inputs:}
  \Statex \hspace*{\algorithmicindent}\parbox[t]{.8\linewidth}{\raggedright #1}
}
\algnewcommand{\Initialize}[1]{%
  \State \textbf{Initialize:}
  \Statex \hspace*{\algorithmicindent}\parbox[t]{.8\linewidth}{\raggedright #1}
}

\algnewcommand{\Returns}[1]{%
  \State \textbf{Return:}
  \Statex \hspace*{\algorithmicindent}\parbox[t]{.8\linewidth}{\raggedright #1}
}

\@ifundefined{showcaptionsetup}{}{%
 \PassOptionsToPackage{caption=false}{subfig}}
\usepackage{subfig}
\makeatother

\providecommand{\lemmaname}{Lemma}
\providecommand{\theoremname}{Theorem}

\begin{document}
\title{Heterogeneous Measurement Selection for Vehicle Tracking using Submodular
Optimization}

\author{%
Matthew R. Kirchner\\ 
Department of Electrical and Computer Engineering\\
University of California, Santa Barbara\\
Santa Barbara, CA 93106-9560\\
kirchner@ucsb.edu
\and 
Jo{\~a}o P. Hespanha\\
Department of Electrical and Computer Engineering\\
University of California, Santa Barbara\\
Santa Barbara, CA 93106-9560\\
hespanha@ece.ucsb.edu
\and 
Denis Garagi{\'c}\\
BAE Systems -- FAST Labs\\
600 District Avenue\\
Burlington, MA 01803\\
denis.garagic@baesystems.com
%%%% IMPORTANT: Use the correct copyright information--IEEE, Crown, or U.S. government. %%%%%
%\thanks{\footnotesize 978-1-7281-2734-7/20/$\$31.00$ \copyright2020 IEEE}              % This creates the copyright info that is the correct 2020 data.
%\thanks{{U.S. Government work not protected by U.S. copyright}}         % Use this copyright notice only if you are employed by the U.S. Government.
%\thanks{{978-1-7281-2734-7/20/$\$31.00$ \copyright2020 Crown}}          % Use this copyright notice only if you are employed by a crown government (e.g., Canada, UK, Australia).
%\thanks{{978-1-7281-2734-7/20/$\$31.00$ \copyright2020 European Union}}    % Use this copyright notice is you are employed by the European Union.
}

\maketitle

\thispagestyle{plain}
\pagestyle{plain}

\begin{abstract}
We study a scenario where a group of agents, each with multiple heterogeneous
sensors are collecting measurements of a vehicle and the measurements
are transmitted over a communication channel to a centralized node
for processing. The communication channel presents an information-transfer
bottleneck as the sensors collect measurements at a much higher rate
than what is feasible to transmit over the communication channel.
In order to minimize the estimation error at the centralized node,
only a carefully selected subset of measurements should be transmitted.
We propose to select measurements based on the Fisher information
matrix (FIM), as ``minimizing'' the inverse of the FIM is required
to achieve small estimation error.

Selecting measurements based on the FIM leads to a combinatorial optimization
problem. However, when the criteria used to select measurements is
both monotone and submodular it allows the use of a greedy algorithm
that is guaranteed to be within $1-1/e\approx63\%$ of the optimum
and has the critical benefit of quadratic computational complexity.
To illustrate this concept, we derive the FIM criterion for different
sensor types to which we apply FIM-based measurement selection. The
criteria considered include the time-of-arrival and Doppler shift
of passively received radio transmissions as well as detected key-points
in camera images.
\end{abstract}

\tableofcontents

%%%%%%%%%%%%%%%%%%%%%%%%%%%%%%%%%%%%%%
\section{Introduction}
%%%%%%%%%%%%%%%%%%%%%%%%%%%%%%%%%%%%%%
\begin{figure}
\begin{centering}
\includegraphics[width=7cm]{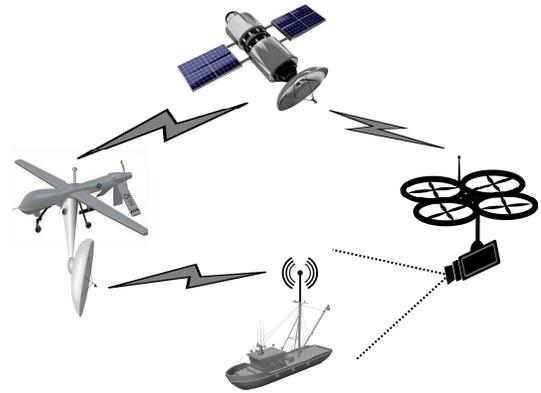}
\par\end{centering}
\caption{An example scenario where two agents separately take measurements
of a ship and transmit them to a satellite for estimation. The ship
being tracked has a radio transmitter and the agent on the left passively
receives the signal. The second agent on the right is observing the
ship with a camera. }
\end{figure}

The scenario we consider is that of a group of agents, each with multiple
sensors collecting noisy measurements of a vehicle, and the measurements
are transmitted over a communication channel to a centralized node.
The central node collects the measurements and estimates a vector
of unknown parameters that describes the motion of the vehicle. The
communication channel restricts the amount of measurements that can
be transmitted to the centralized node. Inspired by the Cram{\'e}r-Rao
lower bound for the error estimate, we propose to select measurements
based on the Fisher information matrix (FIM), as ``minimizing''
the inverse of the FIM is required to achieve small estimation error at the centralized node.

One can use the FIM as a criteria to select which subset of measurements
are ``best'' by formulating a combinatorial optimization problem.
However, this presents a computational challenge as finding the optimal
selection of measurements is, in general, NP-hard. We show that one
common criteria used to ``minimize'' the inverse of the FIM, maximizing
$\log\det\left(\text{FIM}\right)$, is both monotone and submodular
and therefore allows the use of a greedy algorithm \cite[Chapter 16]{cormen2009introduction}
to find the selection of measurements. While the greedy algorithm
returns a sub-optimal solution, it is guaranteed to be within $1-1/e\approx63\%$
of the optimum and has the critical benefit of quadratic computational
complexity.

There have been numerous proposals \cite{shamaiah2010greedy,jawaid2015submodularity,krause2007near,krause2008robust}
to use submodular optimization for sensor selection, however, these
typically seek to optimize a criteria based directly on the estimated
error covariance. As a result, they require simplified estimation
models such as linear Kalman filtering to be used as in \cite{shamaiah2010greedy,jawaid2015submodularity}
or Gaussian process regression on a fixed, discrete grid of points
as in \cite{krause2007near,krause2008robust}. In general, estimation
problems involving vehicle tracking contain non-linear dynamics and
sensor models that result in non-Gaussian and often non-unimodal distributions,
even when all the observation noises are simple independent zero-mean
additive Gaussian distributions.

A key advantage of using the FIM is that we can utilize relatively
simple and well-described distributions of the measurements, without
having to know the (possibly complicated) distributions of the estimation
error. This allows us to decompose the problem into two independent
parts, one of measurement selection and another of performing estimates
based on the selected measurements. The estimation can proceed with
advanced estimation schemes such as non-parametric methods like particle
filtering \cite[Ch. 4.3, pp. 96]{thrun2005probabilistic} or optimization
approaches \cite{shankar2019finite}, among others, without
regard to the measurement selection process.

To illustrate this approach, we derive the FIM for different sensor
types to which we apply measurement selection. This includes the time-of-arrival
and Doppler shift of passively received radio transmissions as well
as detected key-points in camera images. We compare the track estimation
of the vehicle with the FIM selected measurements with that of random
selection and show that selecting measurements based on the FIM can
greatly outperform the estimation task when the bandwidth limitation
becomes significant.

\subsection{Problem Formulation}

Consider an heterogeneous group of $n$ mobile agents, each $i\in\left\{ 1,2,\ldots,n\right\} $
with a sensor that collects a set of $m_{i}$ measurements, which
we denote by
\[
\left\{ y_{k}^{i}:k\in{\mathcal{K}}_{i}\right\} ,
\]
with ${\mathcal{K}}_{i}:=\left\{ 1,2,\ldots,m_{i}\right\} $. Based
on these measurements, we want to estimate a random variable, $\theta$,
of interest at a \emph{measurement fusion center}. Due to bandwidth
limitations, each agent must select a subset of their own measurements
to be transmitted to the remote sensor fusion center. We denote by
$F_{i}\subset{\mathcal{K}}_{i}$ the indices of the measures that agent
$i$ sends to the fusion center. The bandwidth of the wireless channel
imposes a constraint that $\left|F_{i}\right|\leq B_{i}$, where $B_{i}$
is the maximum number of measurements that can be transmitted over
the channel. The set of all measurements available to the fusion center
is given by
\[
{\mathcal{Y}}_{\text{fusion}}=\bigcup_{i=1}^{n}\left\{ y_{k}^{i}:k\in F_{i}\right\} .
\]
Our goal is to select the sets $F_{i}$ such that this set of measurements
contains the ``best'' $\sum_{i}B_{i}$ measurements from the perspective
of estimating $\theta$. Our goal is thus to design algorithms for
each agent so they select subsets $F_{i}$ that optimize
\begin{equation}
\min_{F_{i}\subset{\mathcal{K}}_{i}}\left\{ f\left(F_{i}\right):\left|F_{i}\right|\leq B_{i}\right\} ,\label{eq:Basic problem}
\end{equation}
where $f\left(F_{i}\right)$ is a metric that relates the selected
measurements to estimation performance. We propose to use the Fisher
information matrix (FIM) to construct the functions $f$ as such to
select the best measurements.

Directly finding the optimal value of $\left(\ref{eq:Basic problem}\right)$
is computationally challenging and, in general, NP-hard \cite{bian2006utility}.
This leads to approximations and heuristics to efficiently compute
the selection, such as branch and bound \cite{welch1982branch} or
convex relaxation \cite{joshi2008sensor}. Branch and bound can still
be unreasonably slow while convex relaxations improve speed but is
still cubic in complexity. Neither of these two methods provide any
guarantee on the performance the approximate value relative to the
true optimal.

\section{Fisher Information Matrix}

Assuming that all measurements $y_{k}$ are conditionally independent
given the unknown random variable, $\theta$, the Bayesian Fisher
Information Matrix (FIM) associated with the estimation of $\theta$
is given by
\begin{equation}
\text{FIM}\left(F\right):=Q_{0}+\sum_{k\in F}Q_{k}\label{eq:FIM structure}
\end{equation}
where
\[
F:=\bigcup_{i=1}^{n}F_{i}
\]
denotes the set of all measurements sent to the fusion center,
\begin{equation}
Q_{0}:={\mathbb{E}}\left[\frac{\partial \log p\left(\theta\right)}{\partial\theta}^{\top}\frac{\partial \log p\left(\theta\right)}{\partial\theta}\right],\label{eq:Q0 def}
\end{equation}
denotes the contribution to the FIM due to the a-priori probability
density function (pdf) $p\left(\theta\right)$ of $\theta$.
\begin{equation}
Q_{k}:={\mathbb{E}}\left[\frac{\partial \log p\left(y_{k}|\theta\right)}{\partial\theta}^{\top}\frac{\partial \log p\left(y_{k}|\theta\right)}{\partial\theta}\right],\label{eq:Qk def}
\end{equation}
is the contribution to the FIM due to the measurement $y_{k}$ with conditional
pdf $p\left(y_{k}|\theta\right)$ given $\theta$. In both $\left(\ref{eq:Q0 def}\right)$
and $\left(\ref{eq:Qk def}\right)$, $\frac{\partial\left(\cdot\right)}{\partial\theta}$
denotes the gradient (as a row vector) with respect to the vector
$\theta$. 

The relevance of the FIM to our problem stems from the (Bayesian)
Cram{\'e}r-Rao lower bound, which under the usual regularity assumptions
on the pdfs gives
\[
{\mathbb{E}}\left[\left(\theta-\hat{\theta}\right)\left(\theta-\hat{\theta}\right)^{\top}\right]\geq\text{FIM}\left(F\right)^{-1},
\]
\cite{gill1995applications}, which conceptually means that \textquotedblleft minimizing\textquotedblright{}
the inverse of the FIM is required to achieve a small estimation error.
In this paper, we propose to minimize the criteria
\begin{align}
 & \log\frac{\det\left(\text{FIM}\left(F\right)^{-1}\right)}{\det\left(\text{FIM}\left(\emptyset\right)^{-1}\right)}\nonumber \\
 & =-\log\det\left(\text{FIM}\left(F\right)\right)+\log\det\left(\text{FIM}\left(\emptyset\right)\right)\label{eq:proposed metric}
\end{align}
which essentially corresponds to minimize the volume of the error
ellipsoid, normalized by the volume of the a-priori error ellipsoid.
Alternative criteria include
\begin{equation}
\frac{\text{trace}\left(\text{FIM}\left(F\right)^{-1}\right)}{\text{trace}\left(\text{FIM}\left(\emptyset\right)^{-1}\right)}\label{eq:trace inv metric}
\end{equation}
or
\begin{equation}
\frac{\lambda_{\max}\left(\text{FIM}\left(F\right)^{-1}\right)}{\lambda_{\max}\left(\text{FIM}\left(\emptyset\right)^{-1}\right)},\label{eq:max eig metric}
\end{equation}
which correspond to minimizing the achievable normalized mean-square
estimation error ${\mathbb{E}}[||\theta-\hat{\theta}||^{2}]$ or the
length of the largest axis of the error ellipsoid, respectively. However,
we shall see shortly that $\left(\ref{eq:proposed metric}\right)$
has the desirable property that it leads to a submodular optimization
when the a-priori contribution to the FIM is nonsingular, whereas
$\left(\ref{eq:trace inv metric}\right)$ and $\left(\ref{eq:max eig metric}\right)$
do not share this property \cite{summers2018corrections,jawaid2015submodularity}.

\section{FIM-Based Measurement Selection}

The selection of a set of measurements by agent $i\in\left\{ 1,2,\ldots,n\right\} $
that minimizes the (normalized) volume of the error ellipsoid associated
with its own measurements, subject to communication constraints, can
be formalized as the following maximization:
\begin{equation}
\begin{cases}
\text{maximize} & f\left(F_{i}\right)\\
\text{subject to} & F_{i}\subset{\mathcal{K}}_{i},\\
 & \left|F_{i}\right|\leq B_{i},
\end{cases}\label{eq: primal optimization problem}
\end{equation}
where $f\left(F_{i}\right)$ is the symmetric of $\left(\ref{eq:proposed metric}\right)$
with $\text{FIM}\left(F_{i}\right)$ given by $\left(\ref{eq:FIM structure}\right)$
with $F=F_{i}$, which leads to
\begin{equation}
f\left(F_{i}\right):=\log\det\left(Q_{0}+\sum_{k\in F_{i}}Q_{k}\right)-\log\det\left(Q_{0}\right).\label{eq:proposed f metric}
\end{equation}
We recall that a scalar-valued function $f:2^{\mathcal{K}}\rightarrow{\mathbb{R}}$
that maps subsets of a finite set ${\mathcal{K}}$ to ${\mathbb{R}}$
is called \emph{submodular} if for every $X\subset Y\subset{\mathcal{K}},\,s\in{\mathcal{K}}\setminus Y$
\begin{equation}
f\left(X\cup\left\{ s\right\} \right)-f\left(X\right)\geq f\left(Y\cup\left\{ s\right\} \right)-f\left(Y\right),\label{eq:submodular def}
\end{equation}
\emph{monotone} if
\begin{equation}
X\subseteq Y\implies f\left(X\right)\leq f\left(Y\right),\label{eq:monotone def}
\end{equation}
and \emph{normalized} if
\begin{equation}
f\left(\emptyset\right)=0.\label{eq:normalized def}
\end{equation}

Submodular functions are important for us because of the following
well-known result in combinatorial optimization, which provides an
algorithm to approximate the solution to $\left(\ref{eq: primal optimization problem}\right)$
that has only quadratic complexity on the number of measurements and
provides formal bounds on the performance of the approximation.
\begin{thm}[\cite{nemhauser1978analysis}]
\label{thm:submodolar bound}~When f$\left(\cdot\right)$
is normalized, monotone and submodular, then the Algorithm \ref{alg:Greedy-Optimization-Algorithm.}
returns a set $F_{i}^{*}$ that leads to a criteria $f\left(F_{i}^{*}\right)$
no less than $1-1/e$ of the optimum of $\left(\ref{eq: primal optimization problem}\right)$. 
\end{thm}
\begin{algorithm}
\begin{algorithmic}[1]
\Inputs{$\left\{ \Sigma_k \right\}_{k\in{\mathcal{K}}_i}$,$\left\{ \mu_k \right\}_{k\in{\mathcal{K}}_i}$}
\Initialize{$Q = Q_0$, $F_i=\emptyset$, $c = 0$}
\While{$c<B_i$}
	\State{$j=\underset{k\in {\mathcal{K}}_i\setminus F_i}{\arg\min}\, f\left(F_i \cup \{k\}\right)-f\left(F_i\right)$}
	\State{$Q = Q + \text{FIM}\left(j\right)$}
	\State{$F_i = F_i \cup j$}
	\State{$c = c+1$}
\EndWhile
\Returns{$F_i$}
\end{algorithmic}

\caption{Greedy Optimization Algorithm.\label{alg:Greedy-Optimization-Algorithm.}}
\end{algorithm}

It turns out that the maximization in $\left(\ref{eq: primal optimization problem}\right)$
has submodular structure:
\begin{thm}
\label{thm:criteria proof}Assuming that the a-priori pdf $p\left(\theta\right)$
leads to a positive definite matrix $Q_{0}$ in $\left(\ref{eq:Q0 def}\right)$,
the function $f\left(\cdot\right)$ in $\left(\ref{eq:proposed f metric}\right)$
is normalized, monotone, and submodular.
\end{thm}
To prove Theorem \ref{thm:criteria proof}, we introduce a result
on general functions $f$ of the form
\begin{equation}
f\left(X\right)=g\left(Q_{0}+\sum_{k\in X}Q_{k}\right),\,\forall S\subseteq{\mathcal{K}},\label{eq: set function form}
\end{equation}
where all $Q_{0},Q_{k},\,k\in{\mathcal{S}}$ are $n\times n$ matrices
and $g$ is a function from ${\mathbb{R}}^{n\times n}$ to ${\mathbb{R}}$.
In the sequel, we use the notation
\[
D_{g}\left(A\right):=\left[\frac{\partial g\left(A\right)}{\partial a_{ij}}\right]_{ij}\in{\mathbb{R}}^{n\times n},
\]
which allow us to write
\begin{align}
\frac{dg\left(A\left(\lambda\right)\right)}{d\lambda} & =\sum_{i=1}^{n}\sum_{j=1}^{n}\left[D_{g}\left(A\left(\lambda\right)\right)\right]_{ij}\frac{d\left[A\left(\lambda\right)\right]_{ij}}{d\lambda}\nonumber \\
 & =\text{trace}\left[D_{g}\left(A\left(\lambda\right)\right)^{\top}\frac{dA\left(\lambda\right)}{d\lambda}\right].\label{eq:write Dg}
\end{align}
\begin{lem}
\label{lem:lemma submodular}Assume that $Q_{0}$ is a symmetric positive
definite matrix, that all the $Q_{k},\,k\in{\mathcal{S}}$ are symmetric
positive semidefinite matrices, and that the function $g:{\mathbb{R}}^{n\times n}\rightarrow{\mathbb{R}}$
has the property that for every pair of symmetric positive definite
matrices $A,B$, we have that $D_{g}\left(A\right)$ and $D_{g}\left(B\right)$
are both symmetric positive semidefinite and

\begin{equation}
A\succeq B\implies D_{g}\left(A\right)\preceq D_{g}\left(B\right),\label{eq:Lemma 1}
\end{equation}
then the function $f$ defined by $\left(\ref{eq: set function form}\right)$
is monotone and submodular.
\end{lem}
\begin{proof}
To prove that $f$ is monotone, pick $X\subset Y\subset{\mathcal{S}}$
and define
\[
\delta\left(\lambda\right):=g\left(Q_{0}+\sum_{k\in X}Q_{k}+\lambda\sum_{k\in Y\setminus X}Q_{k}\right),
\]
$\forall\lambda\in\left[0,1\right]$. For this function we have that
\[
\delta\left(0\right)=g\left(Q_{0}+\sum_{k\in X}Q_{k}\right)=f\left(X\right),
\]
and
\[
\delta\left(1\right)=g\left(Q_{0}+\sum_{k\in X}Q_{k}+\sum_{k\in Y\setminus X}Q_{k}\right)=f\left(Y\right),
\]
and, in view of $\left(\ref{eq:write Dg}\right),$
\begin{align}
\frac{d\delta\left(\lambda\right)}{d\lambda} & =\text{trace}\Bigg[D_{g}\left(Q_{0}+\sum_{k\in X}Q_{k}+\lambda\sum_{k\in Y\setminus X}Q_{k}\right)^{\top}\nonumber \\
 & \times\left(\sum_{k\in Y\setminus X}Q_{k}\right)\Bigg].\label{eq:ddelta-dlambda}
\end{align}
Since $Q_{0}$ is positive definite, $Q_{0}+\sum_{k\in X}Q_{k}+\lambda\sum_{k\in Y\setminus X}Q_{k}$
is also positive definite and by assumption
\[
D_{g}\left(Q_{0}+\sum_{k\in X}Q_{k}+\lambda\sum_{k\in Y\setminus X}Q_{k}\right)\succeq0.
\]
Moreover, because the trace of the product of two positive semidefinite
matrices is non-negative, we conclude from $\left(\ref{eq:ddelta-dlambda}\right)$
that $\frac{d\delta\left(\lambda\right)}{d\lambda}\geq0$ and therefore
\[
\delta\left(0\right)=f\left(X\right)\leq\delta\left(1\right)=f\left(Y\right),
\]
from which monotonicity follows.

To prove that $f$ is submodular, we pick $X\subset Y\subset{\mathcal{S}},\,s\in{\mathcal{S}}\setminus Y$
and now define instead
\begin{align*}
\delta\left(\lambda\right) & :=g\left(Q_{0}+Q_{s}+\sum_{k\in X}Q_{k}+\lambda\sum_{k\in Y\setminus X}Q_{k}\right)\\
 & -g\left(Q_{0}+\sum_{k\in X}Q_{k}+\lambda\sum_{k\in Y\setminus X}Q_{k}\right),
\end{align*}
$\forall\lambda\in\left[0,1\right]$. We now have
\begin{align*}
\delta\left(0\right) & =g\left(Q_{0}+Q_{s}+\sum_{k\in X}Q_{k}\right)-g\left(Q_{0}+\sum_{k\in X}Q_{k}\right)\\
 & =f\left(X\cup\left\{ s\right\} \right)-f\left(X\right),
\end{align*}
and
\begin{align*}
\delta\left(1\right) & =g\left(Q_{0}+Q_{s}+\sum_{k\in X}Q_{k}+\sum_{k\in Y\setminus X}Q_{k}\right)\\
 & -g\left(Q_{0}+\sum_{k\in X}Q_{k}+\sum_{k\in Y\setminus X}Q_{k}\right)\\
 & =f\left(Y\cup\left\{ s\right\} \right)-f\left(Y\right)
\end{align*}
and
\begin{align*}
\frac{d\delta\left(\lambda\right)}{d\lambda} & =-\text{trace}\Bigg[\Bigg(D_{g}\left(Q_{0}+\sum_{k\in X}Q_{k}+\lambda\sum_{k\in Y\setminus X}Q_{k}\right)\\
 & -D_{g}\left(Q_{0}+Q_{s}+\sum_{k\in X}Q_{k}+\lambda\sum_{k\in Y\setminus X}Q_{k}\right)\Bigg)^{\top}\\
 & \times\left(\sum_{k\in Y\setminus X}Q_{k}\right)\Bigg].
\end{align*}
Since $Q_{0}\succ0$ and $Q_{s}\succeq0$, we conclude from $\left(\ref{eq:ddelta-dlambda}\right)$
that
\begin{align*}
D_{g}\left(Q_{0}+\sum_{k\in X}Q_{k}+\lambda\sum_{k\in Y\setminus X}Q_{k}\right)\\
-D_{g}\left(Q_{0}+Q_{s}+\sum_{k\in X}Q_{k}+\lambda\sum_{k\in Y\setminus X}Q_{k}\right) & \succeq0
\end{align*}
and therefore
\[
\frac{d\delta\left(\lambda\right)}{d\lambda}\leq0,
\]
because the trace of the product of two positive semidefinite matrices
is non-negative. This shows that
\[
\delta\left(1\right)=f\left(Y\cup\left\{ s\right\} \right)-f\left(Y\right)\leq f\left(X\cup\left\{ s\right\} \right)-f\left(X\right)=\delta\left(0\right),
\]
from which submodularity follows.
\end{proof}

\medskip{}

\begin{proof}[Proof of Theorem \ref{thm:submodolar bound}]The
function in $\left(\ref{eq:proposed f metric}\right)$ is normalized
since for $F_{i}=\emptyset$, the two terms cancel and therefore $f\left(\emptyset\right)=0$.
We prove that this function is monotone and submodular by applying
Lemma \ref{lem:lemma submodular} to the function
\[
g\left(A\right):=\log\det\left(A\right)-\log\det\left(Q_{0}\right),\,\forall A\in{\mathbb{R}}^{n\times n},
\]
for which $\left(\ref{eq: set function form}\right)$ precisely matches
$\left(\ref{eq:proposed f metric}\right)$. For this function $g$,
we have that for any symmetric positive definite matrix $A\in{\mathbb{R}}^{n\times n}$
\[
D_{g}\left(A\right)=A^{-T}>0,
\]
and therefore, for every pair of symmetric positive definite matrices
$A,B\in{\mathbb{R}}^{n\times n}$, we have that
\begin{align*}
A\succeq B & \implies B^{-1/2}AB^{-1/2}\succeq I\\
 & \implies\lambda_{i}\left[B^{-1/2}AB^{-1/2}\right]\geq1\\
 & \implies\lambda_{i}\left[B^{1/2}AB^{1/2}\right]\leq1\\
 & \implies B^{1/2}AB^{1/2}\preceq I\\
 & \implies A^{-1}\preceq B^{-1}.
\end{align*}
This shows that we can indeed apply Lemma \ref{lem:lemma submodular}
and the result follows.

\end{proof}

\section{Motion Models\label{sec:Motion-Models}}

We consider a scenario in which the $n$ mobile agents carry a suite
of onboard sensors to estimate the trajectory of a vehicle and denote
by $q\left(t\right)$ the vehicle's position at time $t$, expressed
in an inertial coordinate system. We consider a constant curvature
motion model for $q\left(t\right)$. Assuming that the vehicle's linear
and angular velocities have constant coordinates $v^{b}$ and $\omega^{b}$,
respectively, when expressed in the body frame, we have
\[
\dot{q}=Rv^{b},
\]
and
\[
\dot{R}=RJ\left(\omega^{b}\right),
\]
with $R\in\text{SO}\left(3\right)$. For this model, the coordinates
of the linear and angular accelerations expressed in the inertial
frame satisfy the equations
\[
\dot{v}^{i}=\dot{R}v^{b}=RJ\left(\omega^{b}\right)v^{b}=J\left(R\omega^{b}\right)Rv^{b}=J\left(\omega^{i}\right)v^{i},
\]
and
\[
\dot{\omega}^{i}=\dot{R}\omega^{b}=RJ\left(\omega^{b}\right)\omega^{b}=0,
\]
which leads to the motion model
\[
\dot{q}=v^{i},
\]
and
\[
\dot{v}^{i}=J\left(\omega^{i}\right)v^{i},
\]
where we can view the angular velocity $\omega^{i}$ as an (unknown)
constant parameter. This differential equation can be integrated exactly
on a time interval $t\in\left[t_{\ell-1},t_{\ell}\right]$ since
\begin{align*}
v^{i}\left(t\right) & =e^{J\left(\omega^{i}\right)\left(t-t_{\ell-1}\right)}v^{i}\left(t_{\ell-1}\right)\\
 & =\Bigg(I+\sin\left(\rho\left(t-t_{\ell-1}\right)\right)J\left(\bar{\omega}\right)\\
 & +\left(1-\cos\left(\rho\left(t-t_{\ell-1}\right)\right)\right)J\left(\bar{\omega}\right)^{2}\Bigg)v^{i}\left(t_{\ell-1}\right),
\end{align*}
where
\[
\bar{\omega}:=\frac{\omega^{i}}{\left\Vert \omega^{i}\right\Vert },
\]
and
\[
\rho:=\left\Vert \omega^{i}\right\Vert .
\]
Since the exact formula for $q\left(t\right)$ is complex, we use
its 2nd order Taylor series approximation for $t$ close to $t_{\ell-1}$,
which leads to
\[
v^{i}\left(t\right)=v^{i}\left(t_{\ell-1}\right)+\left(t-t_{\ell-1}\right)J\left(\omega^{i}\right)v^{i}\left(t_{\ell-1}\right),
\]
and
\begin{align}
q\left(t\right) & =q\left(t_{\ell-1}\right)+\left(t-t_{\ell-1}\right)v^{i}\left(t_{\ell-1}\right)\label{eq: curvature motion model}\\
 & +\frac{\left(t-t_{\ell-1}\right)}{2}J\left(\omega^{i}\right)v^{i}\left(t_{\ell-1}\right).\nonumber 
\end{align}
This motion model can be summarized as
\begin{equation}
q\left(t\right)=\theta_{1}+\left(t-t_{\ell-1}\right)\theta_{2}+\frac{\left(t-t_{\ell-1}\right)^{2}}{2}\theta_{3},\label{eq:motion model}
\end{equation}
where $\theta_{1},\theta_{2},\theta_{3}\in{\mathbb{R}}^{2}$ can be
viewed as three parameters that need to be estimated. These parameters
can be viewed as the target's position, linear velocity, and curvature
on the interval $t\in\left[t_{\ell-1},t_{\ell}\right]$. For targets
moving in along a straight line, this model simplifies to the case
$\theta_{3}=0$ and for stationary targets $\theta_{2}=\theta_{3}=0$.

\section{\label{sec:Measurement-Models}Measurement Models}

We denote by $p\left(t\right)\in{\mathbb{R}}^{3}$ the known position
of a sensing agent that collects measurements to estimate the vehicle's
trajectory. In this section, we specifically consider on-board RF
sensors that measure (i) the times of arrival of radio packets emitted
by the vehicle; (ii) the Doppler frequency shift in their carrier
frequency arising from the relative motion between the vehicle and
receiver; and (iii) image coordinates of distinguishable features
of the vehicle, collected by on-board visible or IR cameras. With
regard to the RF measurements, we do not assume the vehicle's transmissions
are synchronized with the receiver's clock nor precise knowledge of
the vehicle's carrier frequency, which essentially means that our
measurements should be viewed as time difference of arrival (TDoA)
and frequency difference of arrival (FDoA). Because of the TDoA and
FDoA ambiguity, in additional to the motion model parameters $\theta_{1},\theta_{2},\theta_{3}$
in $\left(\ref{eq:motion model}\right)$, we also need to estimate
sensor-specific parameters that account for the lack of synchronization
and knowledge of the carrier frequency.

The remainder of this section, discusses the different sensor measurement
models and implicitly assume that the different sensors produce conditionally
independent measurements, given all the parameters that need to be
estimated. We also assume here that measurements $y_{k}$ occur at
times $\tau_{k}\in\left[t_{\ell-1},t_{\ell}\right]$ and have multi-variable
normal (conditional) distributions with mean $\mu_{k}(\theta)$ that
depends on the vector $\theta$ of unknown parameters and covariance
$\Sigma_{k}$ that, for simplicity, does not depend on $\theta$.
In this case, the matrix $Q_{k}$ in $\left(\ref{eq:Qk def}\right)$
is given by 
\[
Q_{k}=\frac{\partial\mu_{k}\left(\theta\right)}{\partial\theta}^{\top}\Sigma_{k}^{-1}\frac{\partial\mu_{k}\left(\theta\right)}{\partial\theta},
\]
\cite{malago2015information}, where $\frac{\partial\mu_{k}\left(\theta\right)}{\partial\theta}$
denotes the Jacobian matrix of $\mu_{k}$. Because of the nonlinearity
of the map $\theta\mapsto\mu_{k}(\theta)$, the a-posteriori distribution
of $\theta$ given such measurements will typically be non-Gaussian
and often multi-modal.

\subsection{Time-of-Arrival Measurements\label{subsec:Time-of-Arrival-Measurements}}

The vehicle's radio transmitter sends symbols at times $\tau_{k}\in\left[t_{\ell-1},t_{\ell}\right]$
\[
\tau_{k}=kT+T_{0},
\]
with $k\in\left\{ 0,1,\ldots,K-1\right\} $ where $T$ is only approximately
know and $T_{0}$ is unknown to the receiver. Note that $T_{0}$ need
not be the same as the initial time of the estimation interval, $t_{\ell-1}$.
The receiver records noisy observations of the times-of-arrival of
the symbols, denoted by ${\mathcal{T}}$ , which is given by
\[
{\mathcal{T}}\left(\tau_{k}\right)=\tau_{k}+\frac{\rho\left(\tau_{k}\right)}{c},
\]
where $c$ denotes the speed of light and 
\begin{equation}
\rho\left(t\right):=\left\Vert q\left(t\right)-p\left(t\right)\right\Vert .\label{eq:rho def}
\end{equation}
Therefore, the times-of-arrival scaled by the speed of light are given
by
\begin{equation}
c{\mathcal{T}}\left(\tau_{k}\right)=\rho\left(\tau_{k}\right)+\theta_{T}k+\theta_{T_{0}},\label{eq: scaled time of arrival def}
\end{equation}
where
\[
\theta_{T}:=cT,
\]
and
\[
\theta_{T_{0}}:=cT_{0}.
\]
This model assumes that the relative motion between transmitter and
receiver is sufficiently slow so that the receiver\textquoteright s
position at the time the symbol is received is the essentially the
same as when it was transmitted. We regard the noisy measurements
of the times-of-arrival as Gaussian random variables with means given
by the actual times-of-arrival in $\left(\ref{eq: scaled time of arrival def}\right)$
and variance $\sigma_{c{\mathcal{T}}}^{2}$ independent of the unknown
parameters, which means that the likelihood of a measurement $y_{k}$
of $\left(\ref{eq: scaled time of arrival def}\right)$ is given by
\[
P\left(y_{k};\sigma_{c{\mathcal{T}}},\theta_{T},\theta_{T_{0}}\right)=\frac{1}{\sqrt{s\pi}\sigma_{c{\mathcal{T}}}}e^{-\frac{\left(\rho\left(\tau_{k}\right)+\theta_{T}k+\theta_{T_{0}}-y_{k}\right)^{2}}{2\sigma_{c{\mathcal{T}}}}}.
\]
The parameters $\theta_{T}$ and $\theta_{T_{0}}$ are typically not
known a priori and must be estimated jointly along with $\theta_{1}$,
$\theta_{2}$, and $\theta_{3}$. For the motion model in $\left(\ref{eq:motion model}\right)$,
the gradient with respect to the motion-specific parameters is given
by
\[
\frac{\partial c{\mathcal{T}}\left(\tau_{k}\right)}{\partial\theta_{1}}=\frac{\left(q\left(\tau_{k}\right)-p\left(\tau_{k}\right)\right)}{\rho\left(\tau_{k}\right)},
\]
and
\[
\frac{\partial c{\mathcal{T}}\left(\tau_{k}\right)}{\partial\theta_{2}}=\frac{\left(q\left(\tau_{k}\right)-p\left(\tau_{k}\right)\right)}{\rho\left(\tau_{k}\right)}\left(\tau_{k}-t_{\ell-1}\right),
\]
and
\[
\frac{\partial c{\mathcal{T}}\left(\tau_{k}\right)}{\partial\theta_{3}}=\frac{\left(q\left(\tau_{k}\right)-p\left(\tau_{k}\right)\right)}{\rho\left(\tau_{k}\right)}\frac{\left(\tau_{k}-t_{\ell-1}\right)^{2}}{2}.
\]
The gradient with respect to the sensor specific parameters $\theta_{T}$,
and $\theta_{T_{0}}$, is given as
\[
\frac{\partial c{\mathcal{T}}\left(\tau_{k}\right)}{\partial\theta_{T}}=k,
\]
and
\[
\frac{\partial c{\mathcal{T}}\left(\tau_{k}\right)}{\partial\theta_{T_{0}}}=1.
\]

\subsection{\label{subsec:Doppler-Measurements}Doppler Measurements}

The vehicle's radio decoder detects the frequency shift of the received
carrier, which results from both a mismatch between transmitter and
receiver center frequencies as well as from the relative motion between
transmitter and receiver. Specifically, the frequency shift associated
with the $k$th symbol is given by for $\tau_{k}\in\left[t_{\ell-1},t_{\ell}\right]$
\begin{equation}
{\mathcal{F}}\left(\tau_{k}\right)=\Delta f\left(\tau_{k}\right)-\frac{\dot{\rho}\left(\tau_{k}\right)}{\lambda},\label{eq:freq shift}
\end{equation}
where $\lambda=\frac{c}{f_{c}}$, $f_{c}$ is the carrier frequency
of the transmitter, $\Delta f$ is the difference between the carrier
frequencies of the transmitter and receiver, and $\rho$ is defined
in $\left(\ref{eq:rho def}\right)$, leading to
\[
\dot{\rho}\left(t\right)=\frac{\left(q\left(t\right)-p\left(t\right)\right)^{\top}\left(\dot{q}\left(t\right)-\dot{p}\left(t\right)\right)}{\left\Vert q\left(t\right)-p\left(t\right)\right\Vert }.
\]
Therefore, the frequency shifts $\left(\ref{eq:freq shift}\right)$
scaled by the wave length are given by
\begin{equation}
\lambda{\mathcal{F}}\left(\tau_{k}\right)=\theta_{\lambda}-\dot{\rho}\left(\tau_{k}\right),\label{eq: scaled freq shift}
\end{equation}
where
\[
\theta_{\lambda}:=\lambda\Delta f.
\]
We regard the noisy measurements of the frequency shifts as Gaussian
random variables with means given by the actual frequency shifts in
$\left(\ref{eq: scaled freq shift}\right)$ and variance $\sigma_{\lambda{\mathcal{F}}}^{2}$
independent of the unknown parameters, from which follows that the
likelihood of a measurement $y_{k}$ of $\left(\ref{eq: scaled freq shift}\right)$
is given by
\[
P\left(y_{k};\sigma_{\lambda{\mathcal{F}}},\theta_{\lambda}\right)=\frac{1}{\sqrt{2\pi}\sigma_{\lambda{\mathcal{F}}}}e^{-\frac{\left(\theta_{\lambda}-\dot{\rho}\left(\tau_{k}\right)-y_{k}\right)^{2}}{2\sigma_{\lambda{\mathcal{F}}}^{2}}}.
\]
Here, the parameter $\theta_{\lambda}$ is typically not know and
also needs to be estimated. For the motion model in $\left(\ref{eq:motion model}\right)$,
the gradient with respect to the motion parameters is given by
\begin{align*}
\frac{\partial\lambda{\mathcal{F}}\left(\tau_{k}\right)}{\partial\theta_{1}}= & \frac{\left(\dot{q}\left(\tau_{k}\right)-\dot{p}\left(\tau_{k}\right)\right)^{\top}}{\left\Vert q\left(\tau_{k}\right)-p\left(\tau_{k}\right)\right\Vert }\\
 & \times\left(I-\frac{\left(q\left(\tau_{k}\right)-p\left(\tau_{k}\right)\right)\left(q\left(\tau_{k}\right)-p\left(\tau_{k}\right)\right)^{\top}}{\left\Vert q\left(\tau_{k}\right)-p\left(\tau_{k}\right)\right\Vert ^{2}}\right),
\end{align*}
and
\begin{align*}
\frac{\partial\lambda{\mathcal{F}}\left(\tau_{k}\right)}{\partial\theta_{2}} & =\frac{\left(\dot{q}\left(\tau_{k}\right)-\dot{p}\left(\tau_{k}\right)\right)^{\top}}{\left\Vert q\left(\tau_{k}\right)-p\left(\tau_{k}\right)\right\Vert }\\
 & \times\left(I-\frac{\left(q\left(\tau_{k}\right)-p\left(\tau_{k}\right)\right)\left(q\left(\tau_{k}\right)-p\left(\tau_{k}\right)\right)^{\top}}{\left\Vert q\left(\tau_{k}\right)-p\left(\tau_{k}\right)\right\Vert ^{2}}\right)\\
 & \times\left(\tau_{k}-t_{k-1}\right)\\
 & +\frac{\left(q\left(\tau_{k}\right)-p\left(\tau_{k}\right)\right)^{\top}}{\left\Vert q\left(\tau_{k}\right)-p\left(\tau_{k}\right)\right\Vert },
\end{align*}
and
\begin{align*}
\frac{\partial\lambda{\mathcal{F}}\left(\tau_{k}\right)}{\partial\theta_{3}} & =\frac{\left(\dot{q}\left(\tau_{k}\right)-\dot{p}\left(\tau_{k}\right)\right)^{\top}}{\left\Vert q\left(\tau_{k}\right)-p\left(\tau_{k}\right)\right\Vert }\\
 & \times\left(I-\frac{\left(q\left(\tau_{k}\right)-p\left(\tau_{k}\right)\right)\left(q\left(\tau_{k}\right)-p\left(\tau_{k}\right)\right)^{\top}}{\left\Vert q\left(\tau_{k}\right)-p\left(\tau_{k}\right)\right\Vert ^{2}}\right)\\
 & \times\frac{\left(\tau_{k}-t_{k-1}\right)^{2}}{2}\\
 & +\frac{\left(q\left(\tau_{k}\right)-p\left(\tau_{k}\right)\right)^{\top}}{\left\Vert q\left(\tau_{k}\right)-p\left(\tau_{k}\right)\right\Vert }\left(\tau_{k}-t_{k-1}\right).
\end{align*}
The gradient with respect to $\theta_{\lambda}$is given as
\[
\frac{\partial\lambda{\mathcal{F}}\left(t_{k}\right)}{\partial\theta_{\lambda}}=1.
\]

\subsection{Camera Measurements\label{subsec:Camera-Measurements}}

\begin{figure*}
\begin{centering}
\subfloat[10 measurements selected.]{\centering{}\includegraphics[width=5.5cm]{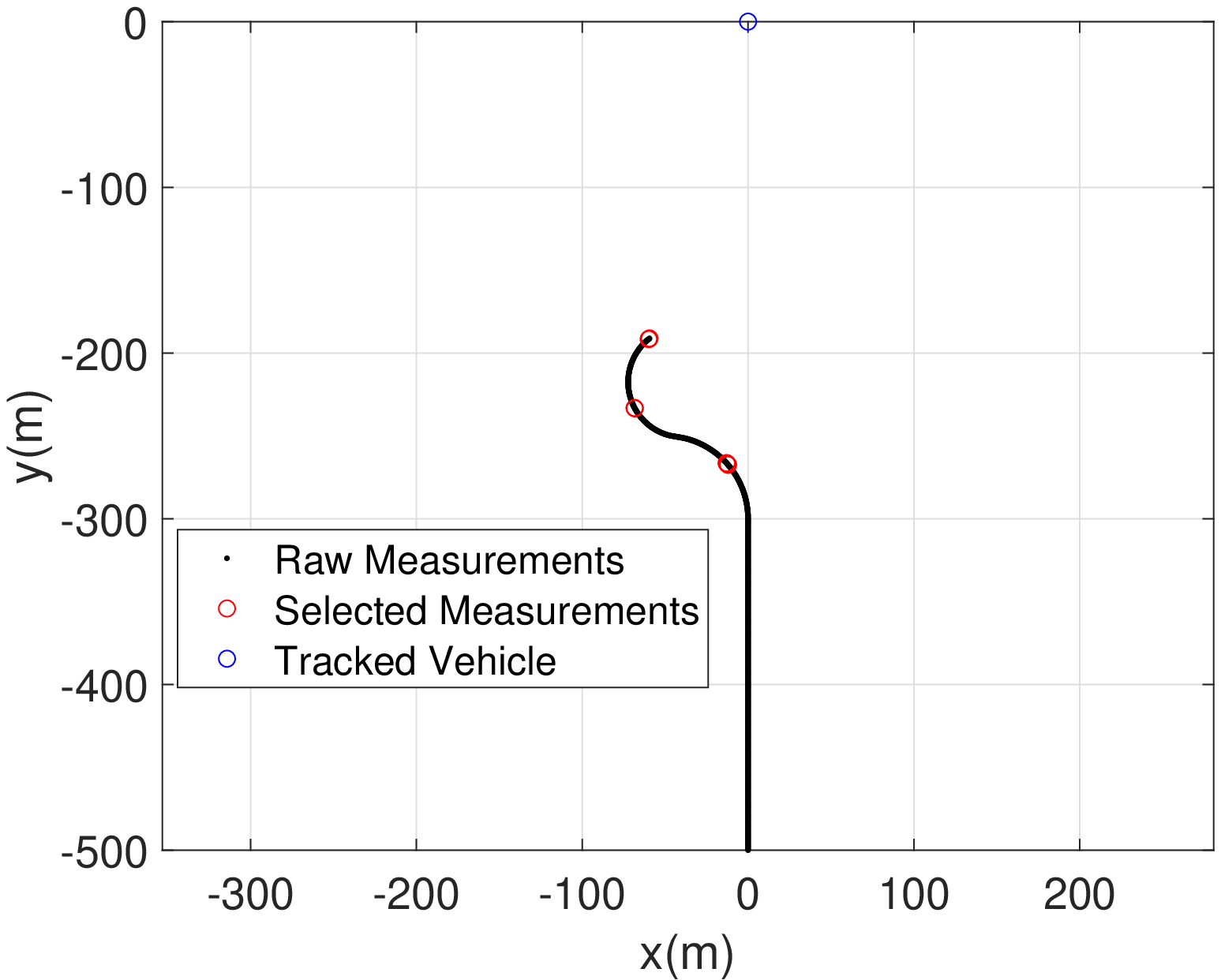}}\hspace*{\fill}\subfloat[50 measurements selected.]{\begin{centering}
\includegraphics[width=5.5cm]{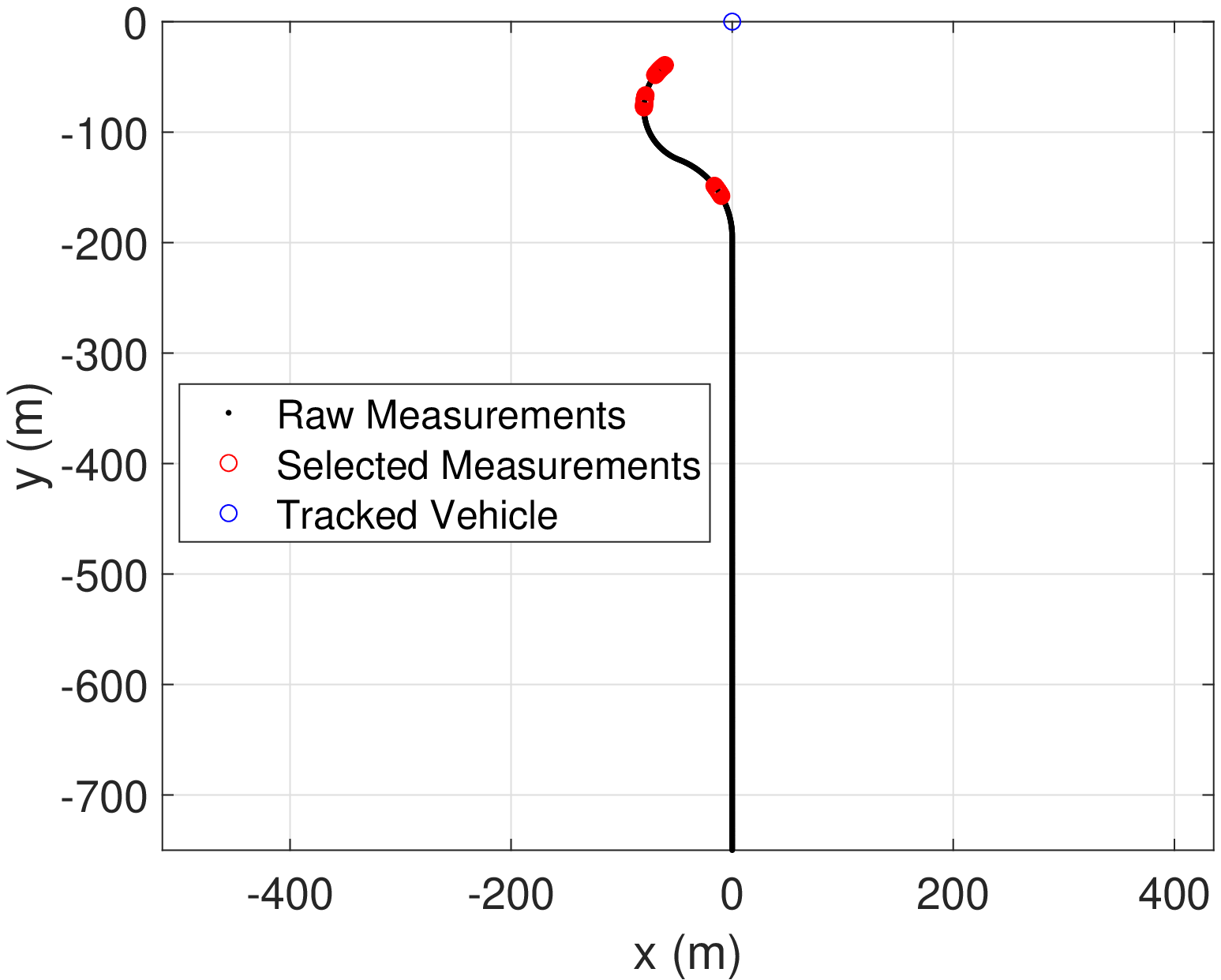}
\par\end{centering}
}\hspace*{\fill}\subfloat[100 measurements selected.]{\centering{}\includegraphics[width=5.5cm]{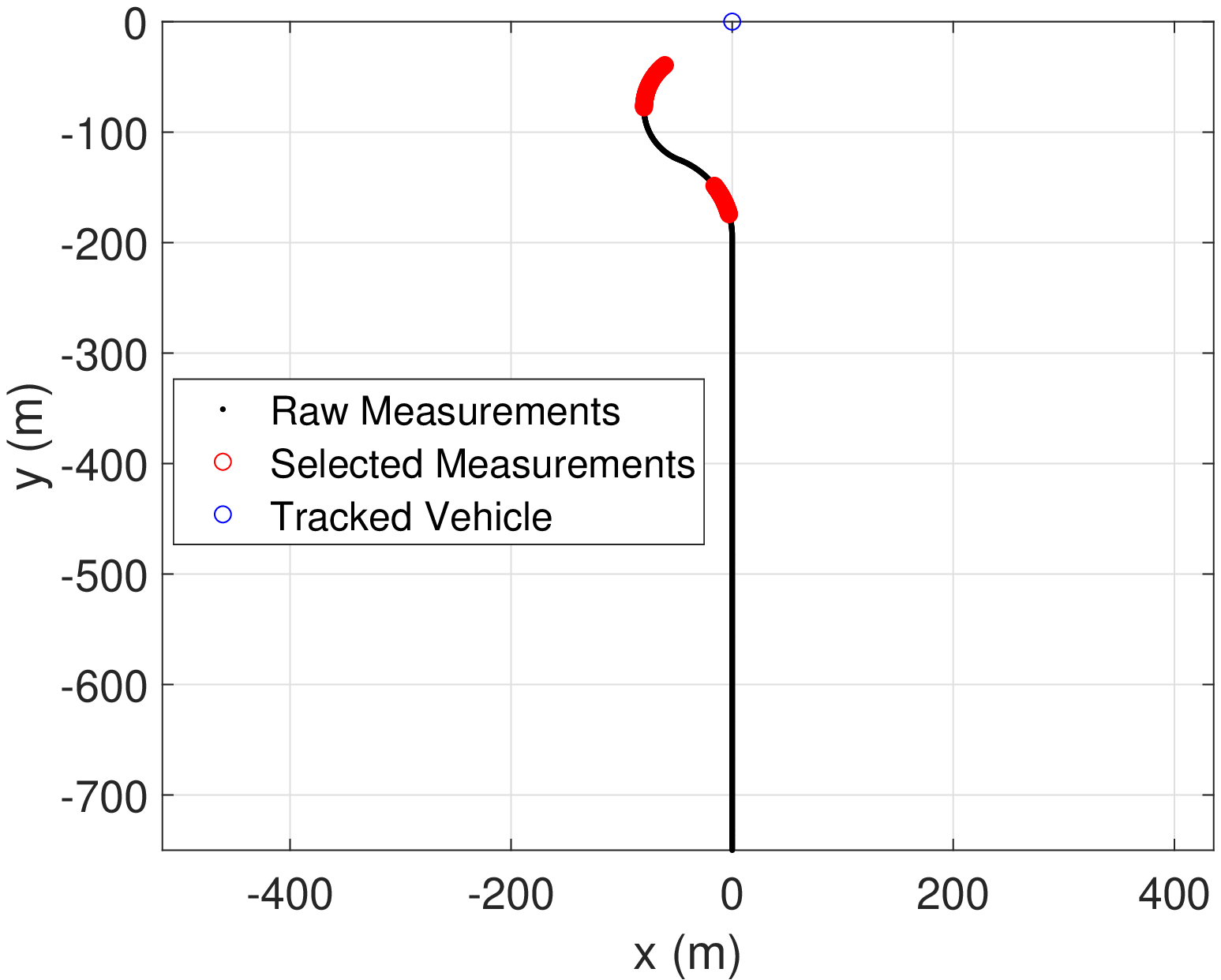}}
\par\end{centering}
\caption{Example 1, where 1000 total measurements are collected uniformly along
the path shown in black. Small subsets of measurements are selected
using the proposed methodology and are shown in red. The ground-truth
location of the object to be tracked is marked in blue.\label{fig:Toy example 1}}
\end{figure*}

The sensing agent has an on-board camera and associated image-processing
algorithms that determine the target's image coordinates. Assuming
a projective camera model with optical center at the agent's position
$p(\tau_{k})$ and a (known) camera orientation $R_{c}(\tau_{k})$,
the target's image coordinates at time $\tau_{k}\in\left[t_{\ell-1},t_{\ell}\right]$
are given as \cite[Chapter. 5, p. 141]{hartley2003multiple}
\begin{equation}
{\mathcal{I}}\left(\tau_{k}\right)=\frac{M\left(\tau_{k}\right)\left(q\left(\tau_{k}\right)-p\left(\tau_{k}\right)\right)}{m\left(\tau_{k}\right)\left(q\left(\tau_{k}\right)-p\left(\tau_{k}\right)\right)},\label{eq:image projection}
\end{equation}
where
\[
M\left(t\right):=\left[\begin{array}{ccc}
1 & 0 & 0\\
0 & 1 & 0
\end{array}\right]AR_{c}\left(t\right),
\]
and
\[
m\left(t\right):=\left[\begin{array}{ccc}
0 & 0 & 1\end{array}\right]AR_{c}\left(t\right),
\]
and $A$ denotes the camera's intrinsic parameters matrix, which is
defined as
\[
A=\left[\begin{array}{ccc}
f_{x} & s_{xy} & o_{x}\\
0 & f_{y} & o_{y}\\
0 & 0 & 1
\end{array}\right],
\]
where $f_{u}$, $f_{v}$ are the camera focal lengths in the $x$,
and $y$ directions, respectively, in the image plane; $\left(o_{x},o_{y}\right)$
is the focal center of the image plane and $s_{xy}$ is the skew parameter.
We regard the target's image coordinates produced by the image processing
algorithms as Gaussian random vectors in ${\mathbb{R}}^{2}$ with means
given by $\left(\ref{eq:image projection}\right)$ and covariance
matrix $\Sigma_{{\mathcal{I}}}$ independent of the unknown parameters,
which means that the likelihood of a measurement $y_{k}$ of $\left(\ref{eq:image projection}\right)$
is given by
\[
P\left(y_{k};\Sigma_{{\mathcal{I}}}\right)=\frac{1}{2\pi\sqrt{\det\Sigma_{{\mathcal{I}}}}}e^{-\frac{1}{2}\left({\mathcal{I}}\left(\tau_{k}\right)-y_{k}\right)^{\top}\Sigma_{{\mathcal{I}}}\left({\mathcal{I}}\left(\tau_{k}\right)-y_{k}\right)}.
\]
For the motion model in $\left(\ref{eq:motion model}\right)$, the
gradient with respect to the motion parameters is given by
\[
\frac{\partial{\mathcal{I}}\left(\tau_{k}\right)}{\partial\theta_{1}}=\frac{M\left(\tau_{k}\right)}{d\left(\tau_{k}\right)}-\frac{U\left(\tau_{k}\right)}{d\left(\tau_{k}\right)^{2}}m\left(\tau_{k}\right),
\]
and
\[
\frac{\partial{\mathcal{I}}\left(\tau_{k}\right)}{\partial\theta_{2}}=\left(\tau_{k}-t_{k-1}\right)\left(\frac{M\left(\tau_{k}\right)}{d\left(\tau_{k}\right)}-\frac{U\left(\tau_{k}\right)}{d\left(\tau_{k}\right)^{2}}m\left(\tau_{k}\right)\right),
\]
and
\[
\frac{\partial{\mathcal{I}}\left(\tau_{k}\right)}{\partial\theta_{3}}=\frac{\left(\tau_{k}-t_{k-1}\right)^{2}}{2}\left(\frac{M\left(\tau_{k}\right)}{d\left(\tau_{k}\right)}-\frac{U\left(\tau_{k}\right)}{d\left(\tau_{k}\right)^{2}}m\left(\tau_{k}\right)\right),
\]
where
\[
U\left(t\right):=M\left(t\right)\left(q\left(t\right)-p\left(t\right)\right),
\]
and
\[
d\left(t\right):=m\left(t\right)\left(q\left(t\right)-p\left(t\right)\right).
\]

\section{Results}

This section contains several examples based on synthetic data that
illustrate the benefits of FIM-based measurement selection.

\subsection{\label{subsec:Example-1:-A}Example 1: A Single Camera}

\begin{figure}
\begin{centering}
\includegraphics[width=7cm]{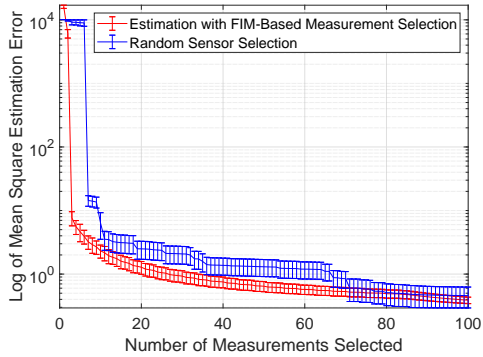}
\par\end{centering}
\caption{\label{fig:Example 1 Est. Error}Averaged estimation error versus
the number of measurements selected for Example 1. Red shows the average
estimation error using measurement selected using the FIM-based method
in Algorithm \ref{alg:Greedy-Optimization-Algorithm.} and blue show
the average estimation error using a random selection.}
\end{figure}

Our first example is that of an agent with a single imaging sensor,
measuring the position of a stationary vehicle in the image plane
as described in Sec. \ref{subsec:Camera-Measurements}. The camera
has a focal length of 50 pixels and image-plane noise with a standard
deviation of $0.8$ pixels. An agent follows the path show in Fig.
\ref{fig:Toy example 1}. This example highlights how the FIM can
be used to asses how informative each of the vision measurements are.
In this instance, when the path is facing straight at the object then
little information is gained in the 'y' direction since any single
image cannot measure depth. However, when multiple images are captured
at different angles relative to the object, then depth can be estimated.
The FIM quantifies this phenomenon and we can optimally pick the most
informative subset of measurements. This is shown in Fig. \ref{fig:Toy example 1}
where 1000 measurements are uniformly collected along the agent's
path and a small subset of measurements are selected using Algorithm
\ref{alg:Greedy-Optimization-Algorithm.} to pick those measurements
that minimally reduce the degradation of estimation performance as
compared to the full set of collected measurements. The FIM-selected
measurements provide a compromise between closeness to the target
(which is maximal right at the end of the path) and largest diversity
of viewing angle. We use the unscented transform \cite{wan2000unscented,julier2002scaled}
to estimate the parameters and the resulting estimation error is shown
in Figure \ref{fig:Example 1 Est. Error}, though other state-of-the-art
estimation schemes designed specifically for tracking could also be
utilized such as interacting multiple model methods \cite{mazor1998interacting}
or changepoint filtering \cite{kirchner2017maneuvering}. Figure \ref{fig:Example 1 Est. Error}
shows tremendous reduction in estimation error when using FIM-based
selection instead of random selection for regimes with very few measurements
selected.

\subsection{\label{subsec:Example-2}Example 2: Two Heterogeneous Sensors}

\begin{figure*}
\begin{centering}
\subfloat[\label{fig:single vehicle-2 sensors trace(inv(fim))}Performance versus
the number of total measurements selected from a pool of 1000 measurements
collected each from a camera and RF sensor.]{\centering{}\includegraphics[width=8.25cm]{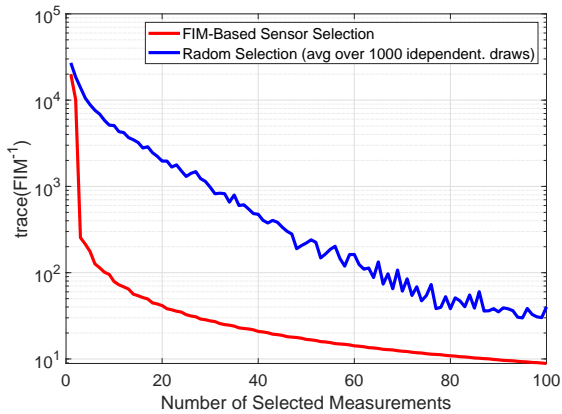}}\hspace*{\fill}\subfloat[\label{fig:Ratio of camera to RF}Ratio of RF and image measurements
selected.]{\begin{centering}
\includegraphics[width=8.25cm]{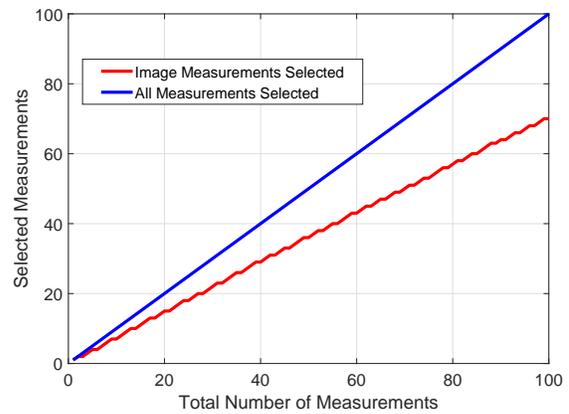}
\par\end{centering}
}
\par\end{centering}
\caption{Example 2, where measurements are jointly selected from two heterogeneous
sensors.}
\end{figure*}

The next example is based on similar trajectories for the vehicle
and mobile agent as in Example 1, but the latter has an additional
RF sensor that measures the Doppler shift, as in Section \ref{subsec:Doppler-Measurements},
with a noise standard deviation of 33ppb in frequency. The agent uses
the FIM-based criteria to select a mixture of measurements between
the two sensors. 
\begin{figure*}
\begin{centering}
\subfloat[10 measurements selected from 1000 measurements collected each from
a camera and RF sensor.]{\centering{}\includegraphics[width=8.25cm]{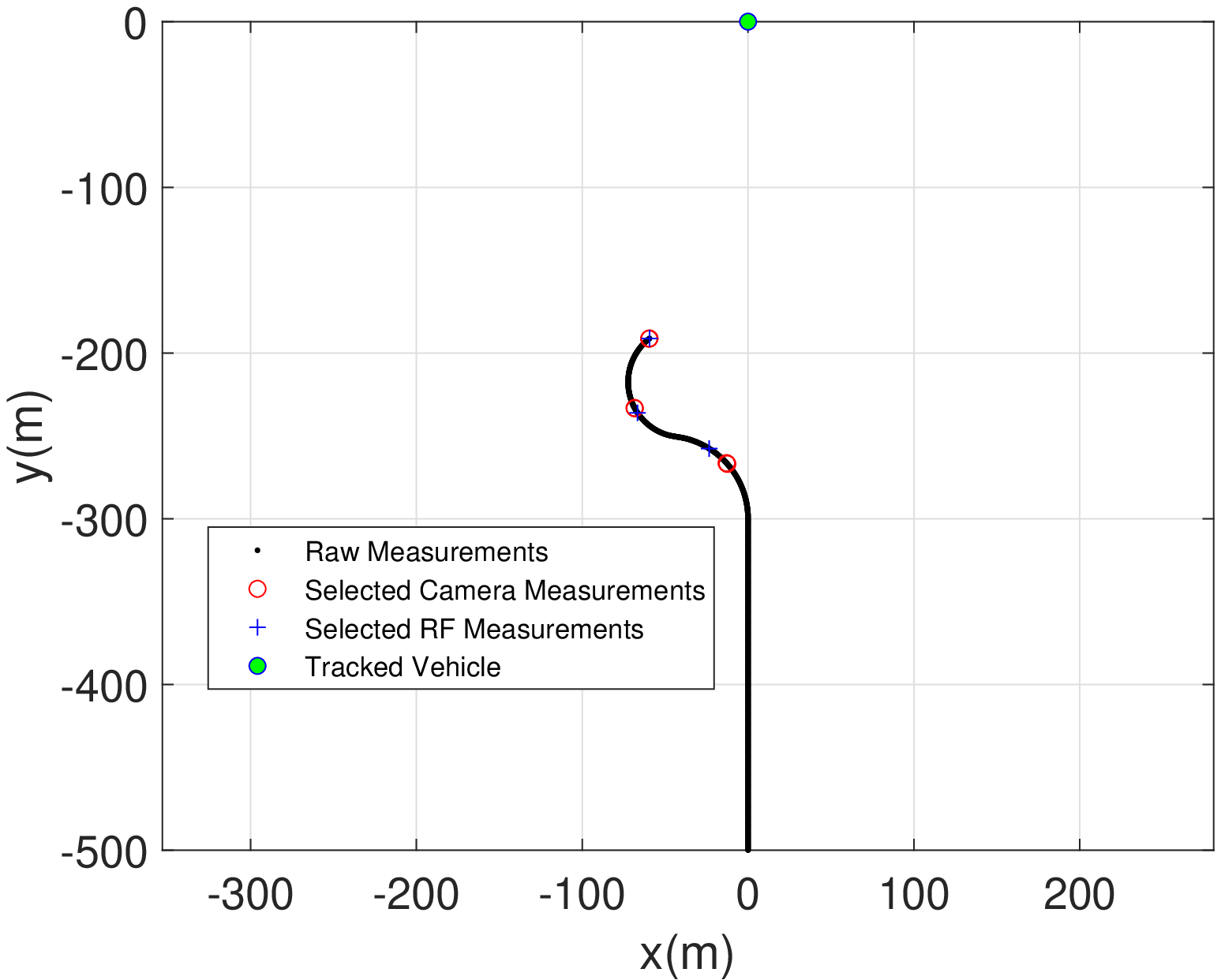}}\hspace*{\fill}\subfloat[Zoomed to show detail of the measurements selected during the final
section of the trajectory.]{\begin{centering}
\includegraphics[width=8.25cm]{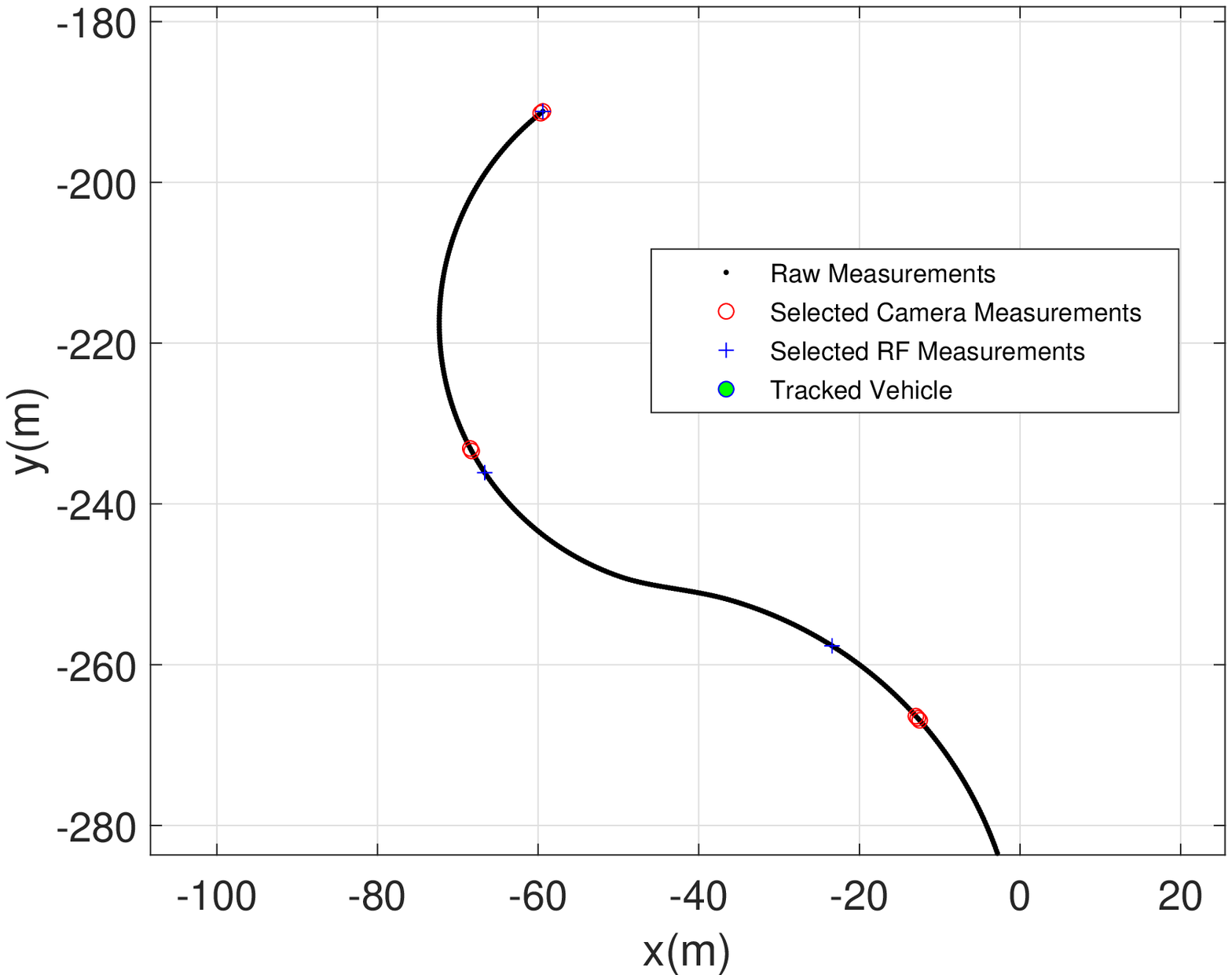}
\par\end{centering}
}
\par\end{centering}
\caption{Example 2, where 1000 total measurements are collected each from a
camera and RF sensor, uniformly along the path shown in black. A subset
of 10 measurements are selected between the sensors using the proposed
methodology. Selected camera measurements are shown red and selected
RF measurements are shown in blue. 7 vision measurements and 3 RF
measurements were selected out of 1000 total available. The ground-truth
location of the object to be tracked is marked in green. }
\end{figure*}
The performance is compared to randomly selecting measurements and
is shown in the Figure \ref{fig:single vehicle-2 sensors trace(inv(fim))},
where we see a dramatic reduction in estimation estimation error with
just a small number of measurements selected. Figure \ref{fig:Ratio of camera to RF}
shows the ratio of image measurements to RF measurements selected.
Where we can see that, for this geometry FIM-based selections roughly
pick 2/3 of the measurements from the camera versus 1/3 from the radio
receiver.

\subsection{\label{subsec:Example-3:-Multiple}Example 3: Multiple Platforms}

\begin{figure}
\begin{centering}
\includegraphics[width=7cm]{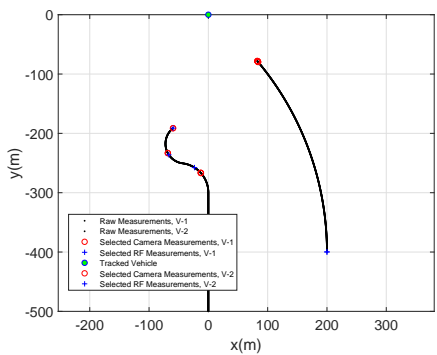}
\par\end{centering}
\caption{\label{fig:2 vehicle selection}Example 3, where each of the 2 agents
collects 1000 total measurements each from a camera and RF sensor,
uniformly along the paths shown in black. A subset of 10 measurements
are selected between the sensors using the proposed methodology from
each agent. Selected camera measurements are shown red and selected
RF measurements are shown in blue. The ground-truth location of the
object to be tracked is marked in green. }
\end{figure}

We expand on Example 2 by having two agents, each with a camera and
RF sensor measuring frequency shifts. Each agent selects a mixture
of camera and RF measurements based in Algorithm \ref{alg:Greedy-Optimization-Algorithm.}
and sends it to a centralized node for processing. Figure \ref{fig:2 vehicle selection}
shows which measurements are selected from each agent. The error covariance
versus the number of measurements selected by each agent is shown
in Figure \ref{fig:Estimation-error-performance / coop}. Compared
to estimation with only a single agent, multiple independent agents
performing FIM-based measurement selection performs better. 

\subsection{\label{subsec:Cooperative-Measurement-Selectio}Cooperative Measurement
Selection}

The measurement selection algorithm provided by Algorithm \ref{alg:Greedy-Optimization-Algorithm.}
operates independently across agents and therefore does not require
inter-agent communication. However, when communication between agents
is available, there is opportunity for further performance improvement,
even when only a small amount of information can be exchanged between
agents.

\begin{figure}
\begin{centering}
\includegraphics[width=7cm]{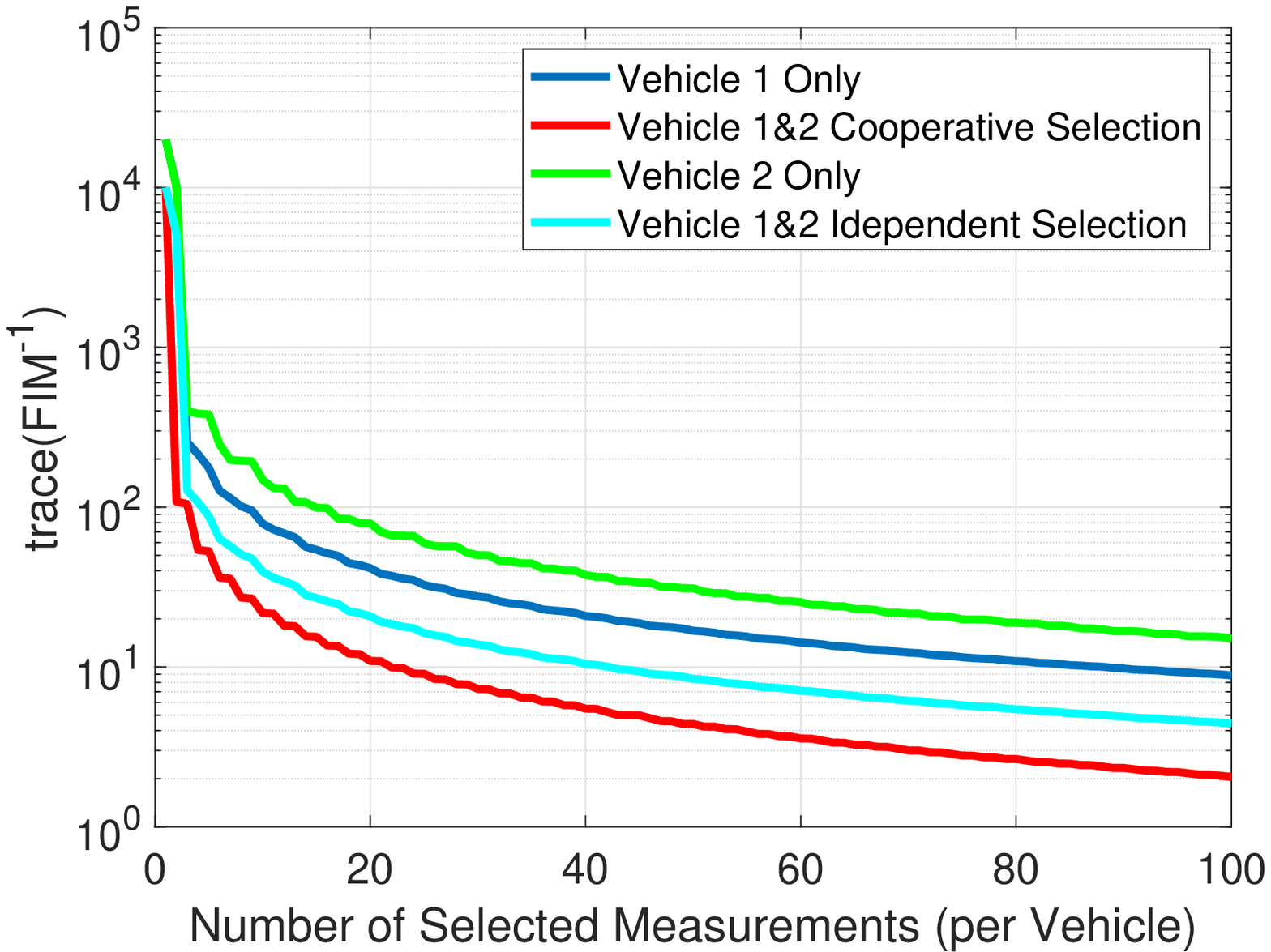}
\par\end{centering}
\caption{\label{fig:Estimation-error-performance / coop}Estimation error performance
using only the given number of measurements selected using the FIM-based
approached. The blue and green show the performance for using only
agent 1 or 2, respectively. The cyan curve is the performance when
both are used at the centralized node to improve estimation after
being selected independently with a FIM-based criteria. Red is the
performance when cooperative FIM-based measurement selection is performed. }
\end{figure}

Consider Example 3 above, but after agent 1 chooses a set of measurements,
the FIM matrix $\text{FIM}\left(F_{i}^{*}\right)$ is shared with
agent 2 to use as a replacement of $Q_{0}$ in its local selection
of measurements, $F_{2}^{*}$. This allows the consideration of agent
1's selection to aid in selecting a better subset from agent 2. The
result is a substantial decrease in estimation error compared to the
independent selection and is shown in Figure \ref{fig:Estimation-error-performance / coop}.

\section{Conclusions and Future Work}

We presented a FIM-based submodular criteria to select measurements
for near-optimal estimation performance in a computationally efficient
manor. We construct the FIM for several sensors that are commonly
used in vehicle tracking problems. Future work includes establishing
theoretical guarantees for cooperative sensor selection and validation
on experimental data. We have preliminary results showing that the
cooperative algorithm outlined in Section \ref{subsec:Cooperative-Measurement-Selectio}
provides theoretical guarantees of performance when compared to the
optimal centralized algorithm.

%%%%%%%%%%%%%%%%%%%%%%%%%%%%%%%%%%%%%%%%%%%%%%%%%%%%%%%%%%%%%%%%%%%%%%%%%%%%%%%%%%%%%%%%%%%%%%%%%%%%%%
\acknowledgments
This work was supported as part of the CogDeCon program funded in part under contract number FA8750-18-C-0014 and in part under contract 88ABW-2019-4739.

\balance

%%%%%%%%%%%%%%%%%%%%%%%%%%%%%%%%%%%%%%%%%%%%%%%%%%%%%%%%%%%%%%%%%%%%%%%%%%%%%%%%%%%%%%%%%%%%%%%%%%%%%%
\bibliographystyle{IEEEtran}
\bibliography{IEEE_Aero_2020}

%%%%%%%%%%%%%%%%%%%%%%%%%%%%%%%%%%%%%%%%%%%%%%%%%%%%%%%%%%%%%%%%%%%%%%%%%%%%%%%%%%%%%%%%%%%%%%%%%%%%%%
\thebiography
%% This biostyle allows you to insert your photo size 1in X 1.25in
\begin{biographywithpic}
{Matthew R. Kirchner}{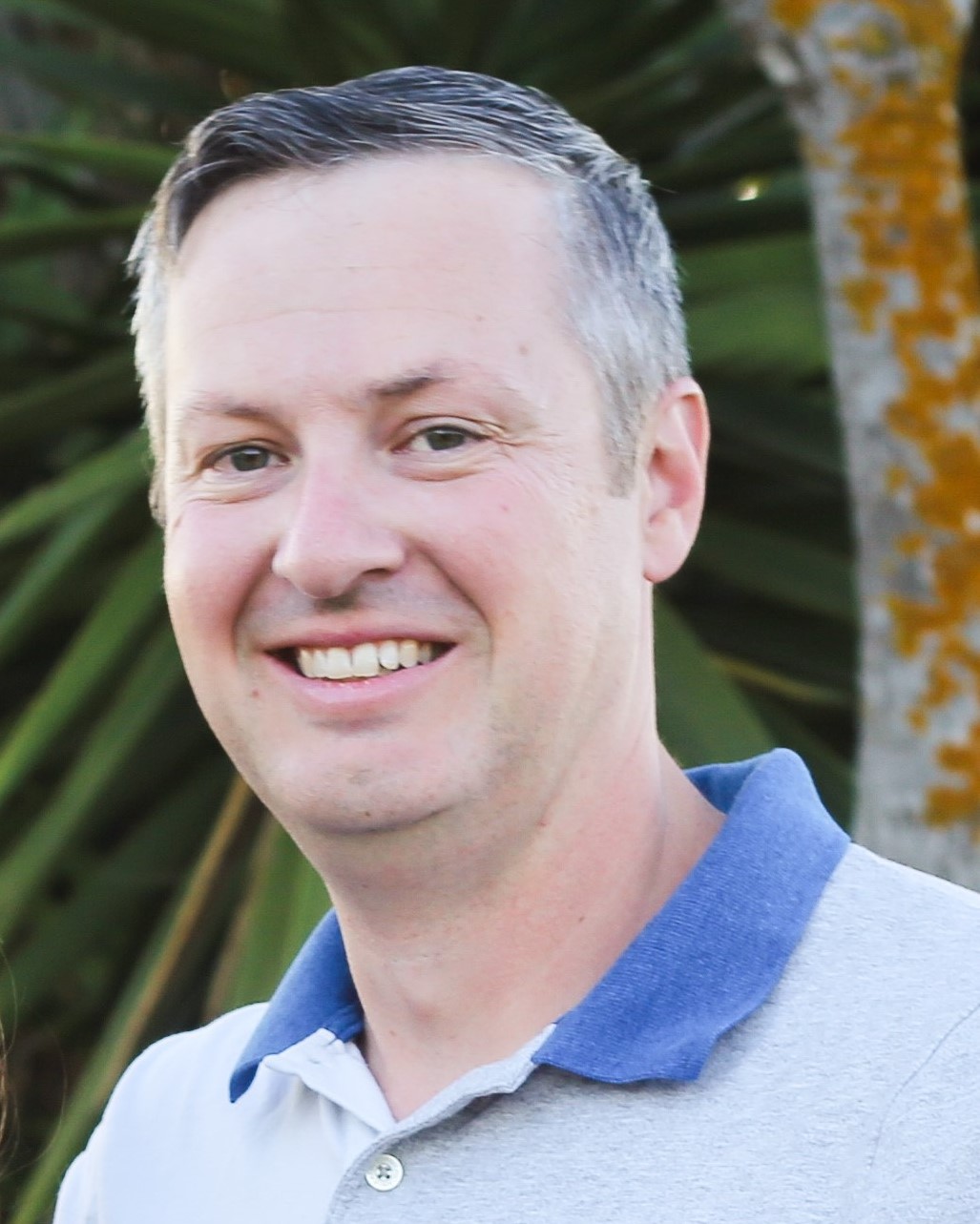}
received his B.S. in Mechanical Engineering from Washington State University in
2007 and his M.S. in Electrical Engineering from the University of Colorado at Boulder in 2013. He joined
the Naval Air Warfare Center Weapons Division in 2007 and since 2012 has been with the Image and
Signal Processing Branch in the Research Directorate, Code 4F0000D. He is currently a Ph.D. student
in the Electrical and Computer Engineering Department at the University of California, Santa Barbara. His
research interests include level set methods for optimal control, differential games, and reachability; multi-vehicle robotics; nonparametric signal and image processing; and navigation and flight control. He was the recipient of a Naval Air
Warfare Center Weapons Division Graduate Academic Fellowship from 2010 to 2012 and in 2011 was
named a Paul Harris Fellow by Rotary International. Matthew is a student member of the IEEE.
\end{biographywithpic} 

\begin{biographywithpic}
{Jo{\~a}o P. Hespanha}{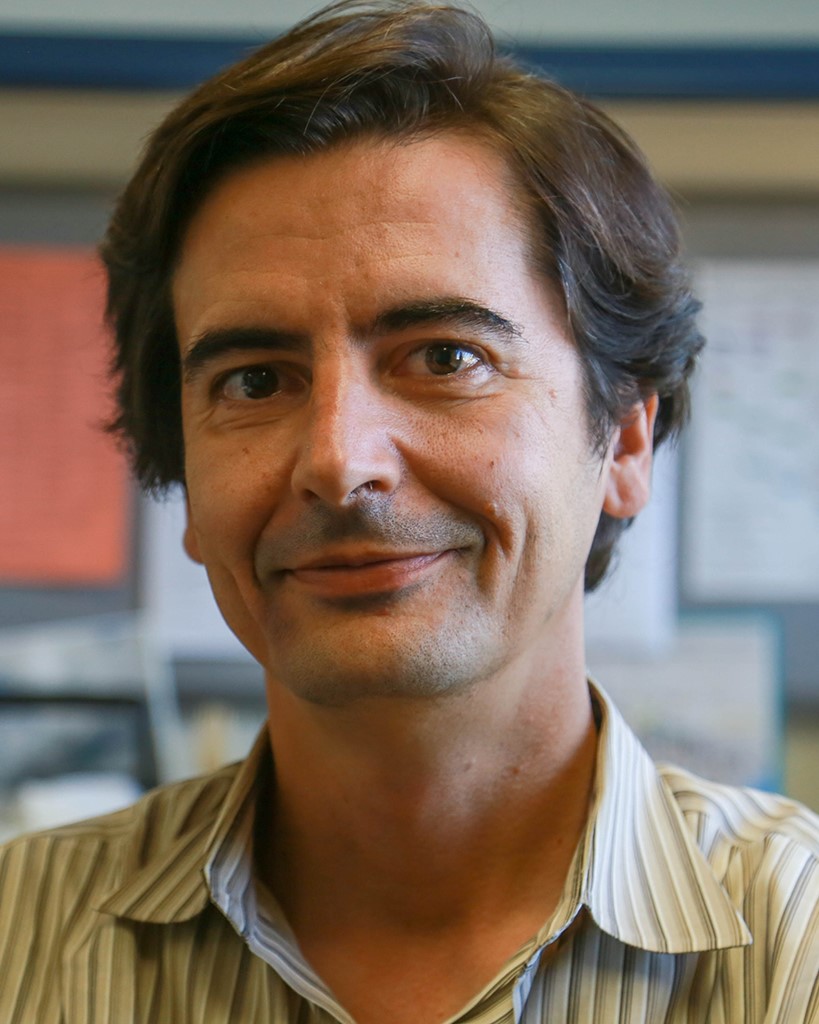}
received his Ph.D. degree in electrical engineering and applied science from Yale University, New Haven, Connecticut in 1998. From 1999 to 2001, he was Assistant Professor at the University of Southern California, Los Angeles. He moved to the University of California, Santa Barbara in 2002, where he currently holds a Professor position with the Department of Electrical and Computer Engineering. Dr. Hespanha is the recipient of the Yale University's Henry Prentiss Becton Graduate Prize for exceptional achievement in research in Engineering and Applied Science, the 2005 Automatica Theory/Methodology best paper prize, the 2006 George S. Axelby Outstanding Paper Award, and the 2009 Ruberti Young Researcher Prize. Dr. Hespanha is a Fellow of the IEEE and he was an IEEE distinguished lecturer from 2007 to 2013. His current research interests include hybrid and switched systems; multi-agent control systems; distributed control over communication networks (also known as networked control systems); the use of vision in feedback control; stochastic modeling in biology; and network security.
\end{biographywithpic}

\begin{biographywithpic}
{Denis Garagi{\'c}}{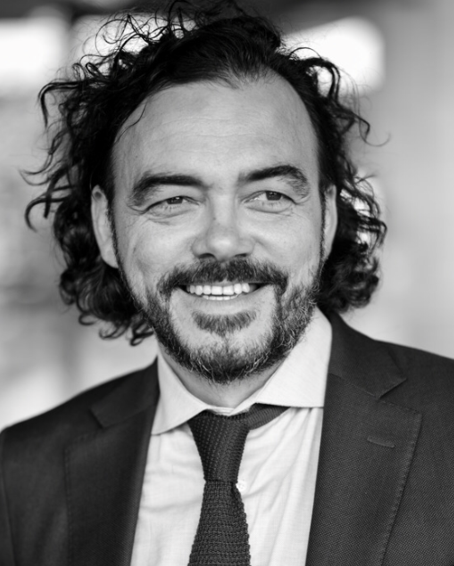}
is a Chief Scientist at BAE Systems FAST Labs. He is a key innovator, guiding FAST Labs's creation of cognitive computing solutions that provide machine intelligence and anticipatory intelligence to solve challenges across any domain for multiple United States Department of Defense customers including DARPA, the services, and the intelligence community. Denis has 20 years of experience in the areas of autonomous cooperative control for unmanned vehicles; game theory for distributed and hierarchical multilevel decision-making, agent-based modeling and simulation; artificial intelligence and machine learning for multi-sensor data fusion, complex scene understanding, motion activity pattern learning and prediction, learning communications signal behavior, speech recognition, and automated text generation. He received his B.S. and M.S. (Mechanical Engineering and Technical Cybernetics; Applied Mathematics) degrees from The Czech Technical University, Prague, Czech Republic, and his Ph.D. in Mechanical Engineering -- System Dynamics and Controls from The Ohio State University.
\end{biographywithpic}

\end{document}